%% file: main_revise2.tex
\documentclass[prl,reprint,superscriptaddress,floatfix,amsmath,amssymb]{revtex4-2}
\usepackage{color}
\usepackage{xcolor}
\usepackage{graphicx}
\usepackage{float}
\usepackage{tikz}
\usepackage{hyperref}
\usepackage[maxfloats=256]{morefloats}
\usepackage{comment}
\usepackage[small, bf]{caption}
\usepackage[subrefformat=parens]{subcaption}
\usepackage{xspace}
\usepackage[version=3]{mhchem}
\usepackage{mathrsfs}
\usepackage{bm}
\usepackage{physics}

\hypersetup{colorlinks=true,urlcolor=blue,citecolor=blue,linkcolor=blue,breaklinks=true}

\usepackage{physrevsupplement}
\usepackage[ruled, linesnumbered]{algorithm2e}

\definecolor{darkgreen}{rgb}{0.0, 0.5, 0.0}

\newcommand{\im}{{\rm i}}
\newcommand{\iw}{\ensuremath{\im \omega}}
\newcommand{\iv}{\ensuremath{\im \nu}}

\newcommand{\wmax}{\ensuremath{{\omega_\mathrm{max}}}}
\usepackage[normalem]{ulem}

\usepackage{scalefnt} 
\usepackage{mathrsfs} 
\newcommand{\cR}{{{\mbox{\scalefont{0.97}$\mathcal{R}$}}}}
\newcommand{\scR}{{{\mbox{\scalefont{0.7}$\mathcal{R}$}}}} 
\newcommand{\sscR}{{\scalefont{0.5}\cR}}
\newcommand{\cL}{{{\mbox{\scalefont{0.97}$\mathcal{L}$}}}}
\newcommand{\scL}{{{\mbox{\scalefont{0.7}$\mathcal{L}$}}}} 
\newcommand{\sscL}{{\scalefont{0.5}\cL}}

\DeclareMathOperator*{\argmax}{argmax}

\newcommand*{\argmaxl}{\argmax\limits}

\usepackage[whole]{bxcjkjatype}
\definecolor{darkgreen}{rgb}{0,0.5,0}
\newcommand{\changed}[1]{{\textcolor{black}{#1}}}

\SetKw{KwEnd}{end}
\SetKw{KwFrom}{from}
\SetKw{KwBreak}{break}
\SetKwFor{For}{for}{do}{\KwEnd}
\SetKwIF{If}{ElseIf}{Else}{if}{then}{else if}{else}{\KwEnd}

\newcommand{\onesVector}{
\noindent
\begin{tikzpicture}[scale=0.2]
    \filldraw (0,0)circle(0.5);
    \draw [very thick](0,0)--(0,1.2);
\end{tikzpicture}
}

\newcommand{\halfVector}{
\noindent
\begin{tikzpicture}[scale=0.2]
    \draw [very thick](0,0)--(0,1.2);
    \filldraw [very thick, white](0.5,0.5)--(0.5,-0.5)--(-0.5,-0.5)--(-0.5,0.5)--cycle;
    \draw [very thick](0.5,0.5)--(0.5,-0.5)--(-0.5,-0.5)--(-0.5,0.5)--cycle;
\end{tikzpicture}
}

\begin{document}

\allowdisplaybreaks
\title{
Low-rank quantics tensor train representations of Feynman diagrams for \\ multiorbital electron-phonon models
}
\author{Hirone Ishida}
\affiliation{Department of Physics, Saitama University, Saitama 338-8570, Japan}

\author{Natsuki Okada}
\affiliation{Department of Physics, Saitama University, Saitama 338-8570, Japan}

\author{Shintaro Hoshino}
\affiliation{Department of Physics, Saitama University, Saitama 338-8570, Japan}

\author{Hiroshi Shinaoka}
\affiliation{Department of Physics, Saitama University, Saitama 338-8570, Japan}

\begin{abstract}
Feynman diagrams are an essential tool for simulating strongly correlated electron systems. However, stochastic quantum Monte Carlo sampling suffers from the sign problem, particularly when solving a multiorbital quantum impurity model. Recently, two approaches have been proposed for efficient numerical treatment of Feynman diagrams: Tensor Cross Interpolation (TCI) to replace stochastic sampling and the Quantics Tensor Train (QTT) representation for compressing space-time dependence.
One of the remaining challenges is the nontrivial task of identifying low-rank structures in weak-coupling Feynman diagrams for multiorbital electron-phonon systems. In particular, the traditional TCI algorithm faces an ergodicity problem, which prevents it from fully exploring the multiorbital space. To address this, we incorporate a new algorithm called global search, which resolves this issue. By combining this approach with QTT, we uncover low-rank structures and achieve efficient numerical integration with exponential resolution in time and faster-than-power-law convergence of error relative to computational cost.
\changed{Additionally, our approach does not require the division of discontinuous regions necessary in non-quantics TCI.}
\end{abstract}

\maketitle
In diverse fields ranging from condensed matter physics to high-energy physics, calculations based on Feynman diagrams in quantum field theory have become indispensable~\cite{Mahan}.
Numerical perturbative calculations often involve numerical integrations and summations over virtual times and internal degrees of freedom of these diagrams. However, when dealing with multiorbital models or higher-order diagrams, straightforward numerical integration faces the notorious ``curse of dimensionality,'' leading to exorbitant computational times. Consequently, methods to circumvent this curse in quantum field theory calculations are in high demand. 

Various approaches have been explored to address this challenge. For instance, quantum Monte Carlo (QMC) methods~\cite{hirsch1986monte, van2010diagrammatic, prokof2007bold, gull2011continuous}, grounded in importance sampling in multidimensional spaces, are widely employed.
Yet, in complex systems, these methods suffer from the negative sign problem, causing computational times to explode exponentially~\cite{shinaoka2015negative, Pan_2024, hann2017solution, zhang2022fermion, mondaini2022quantum}.
This occasionally hinders accessing parameter regimes with physically interesting phenomena.

\changed{Tensor Cross Interpolation (TCI) is a recent advancement using tensor networks to efficiently integrate functions in multidimensional spaces~\cite{fernandez2022learning}. TCI leverages low-rank structures to achieve high precision, surpassing Monte Carlo methods in some cases, such as single-orbital quantum impurity models~\cite{fernandez2022learning,Erpenbeck2023-dz}. It offers faster convergence than QMC, with an error rate of $O(1/n_{\mathrm{s}}^a)$, where $a > 1/2$. However, the factors affecting this rate are still unclear. While TCI has shown promise in simpler systems like double quantum dots~\cite{fernandez2022learning}, its extension to more complex multiorbital systems and electron-phonon interactions is necessary for broader applications.}

\begin{figure}[H]
    \centering
    \includegraphics[width=1.0\linewidth]{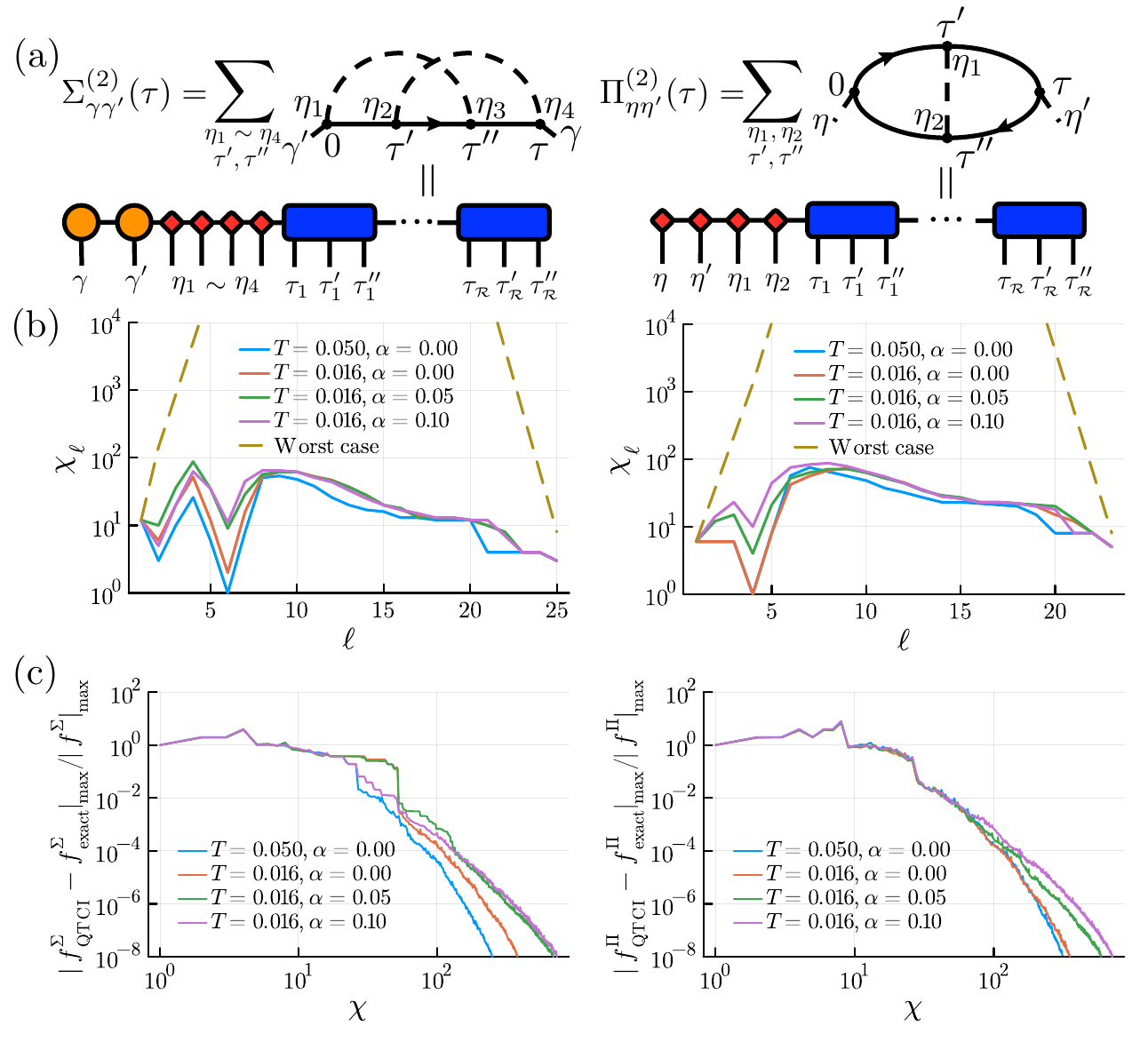}
    \caption{
    (a) Second-order skeleton diagrams for the electron self-energy $\Sigma^{(2)}$ and phonon self-energy $\Pi^{(2)}$, as well as tensor train (TT) representations of their integrands.
    (b) 
    Dimension of the $\ell$-th bond $\chi_\ell$ for TCI tolerance $\epsilon=10^{-3}$ for typical parameters set above/below the superconducting temperature $T_\mathrm{c}\simeq 0.02$ and with/without an external field, whose strength is denoted by $\alpha$.
    We employed a virtual $2^\scR\times 2^\scR \times 2^\scR$ grid for $\tau$, $\tau'$ and $\tau''$ ($\cR=20$, $2^\scR\simeq 10^6$).
    (c) Interpolation error estimate as a function of bond dimension $\chi$. The error is normalized by an estimate of the absolute maximum of the function (integrand).
    }
    \label{fig:highlight}
\end{figure}

\changed{In the meantime, the quantics tensor train (QTT) method has been introduced to solve complex equations like Navier-Stokes and Vlasov-Poisson~\cite{gourianov2022quantum, Ye2022-fu}. It enhances resolution by leveraging low-entanglement structures across different scales~\cite{shinaoka2023multiscale}. Combining TCI and QTT forms quantics tensor cross interpolation (QTCI), useful in condensed matter physics for tasks like Brillouin zone integrals~\cite{ritter2024quantics}. Exploring QTCI for more complex integrations remains a promising area.}

The primary objective of this study is to address the following questions: (i) Do low-rank tensor train (TT) structures exist within self-energy Feynman diagrams for \textit{multiorbital systems}? (ii) Is the TCI robust for interpolating complex integrands with continuous and discrete degrees of freedom? (iii) Can TCI accommodate phonon contributions, which, despite their physical relevance, occasionally worsens the sign problem in QMC simulations? (iv) Does employing quantics facilitate achieving an enhanced convergence rate, specifically \changed{faster-than-power-law} convergence?
To investigate these questions, as a first step, we analyze weak-coupling self-energy Feynman diagrams in the imaginary-time formalism for a prototype multiorbital electron-phonon model, originally proposed for exploring superconducting states in fullerides~\cite{takabayashi2016unconventional, nomura2016exotic,  capone2009colloquium, gunnarsson1997superconductivity, kaga2022eliashberg}.

Figure~\ref{fig:highlight} summarizes our main results obtained with and without an external field.
Figure~\ref{fig:highlight}(a) illustrates the TT structures proposed in this study for the electron and phonon self-energies, where imaginary-time variables are discretized by the quantics representation.
Discrete internal and external variables, such as spin-orbital indices $\gamma$ and vibrational modes of phonon $\eta$ are encoded 
in the same TT as well.
As shown in Fig.~\ref{fig:highlight}(b), the bond dimensions of the TT representations obtained by TCI are much smaller than the worst case of incompressible functions, establishing the existence of low-rank structures.
As demonstrated in Fig.~\ref{fig:highlight}(c), the interpolation error estimate in the TCI vanishes exponentially regardless of the superconducting transition and the presence of the external field, indicating the advantage of the QTT representation.

The Hamiltonian of the model reads
\begin{align}
    \mathscr{H} &=\sum_{ij}\sum_{\gamma\gamma'\sigma}\left(t^{\gamma\gamma'}_{ij} - \mu\delta_{ij}\delta_{\gamma\gamma'}\right)c^\dag_{i\gamma\sigma}c_{j\gamma'\sigma} + \sum_{i\eta}\omega_\eta a^\dag_{i\eta}a_{i\eta}\nonumber\\
    &\quad + \sum_{i\eta}I_\eta :T_{i\eta}T_{i\eta}: + \sum_{i\eta}g_\eta\phi_{i\eta}T_{i\eta} + \alpha\mathscr{H}_{\mathrm{ex}},
\end{align}
where $\gamma=x,y,z$ represent the $t_{1u}$ electronic orbitals, $\eta = 0,1,3,4,6,8$ correspond to the six vibrational modes of a fullerene molecule with energies $\omega_\eta$, and $c^\dag$ and $c$ are the creation and annihilation operators for electrons, respectively, while $a^\dag$ and $a$ are the creation and annihilation operators for phonons.
The first term represents inter-site hopping $t_{ij}$ and the chemical potential $\mu$.
The inter-site hopping is characterized by the semicircular non-interacting density of states with a width of 1.
The second term gives the single-particle phonon energies \(\omega_\eta\).
The third term represents instantaneous local Coulomb interactions.
Here, the ($:$) symbol represents the normal ordering, and $T_{i\eta} = \sum_{\gamma\gamma'\sigma}c^\dag_{i\gamma\sigma}\lambda^\eta_{\gamma\gamma'}c_{i\gamma'\sigma}$,
$I_\eta = (3/4)U-J$ ($\eta = 0$), $=J/2$ (otherwise).
The \(\lambda^\eta\) denote the Gell-Mann matrices with \(\lambda^0\) being defined as the identity matrix, representing the isotropic vibrations of the fullerene molecule. The other values of \(\eta\) correspond to the standard definition of the Gell-Mann matrices and represent anisotropic vibrations.
The fourth term denotes the electron-phonon couplings, with $\phi_{i\eta} \equiv a_{i\eta} + a^\dag_{i\eta}$.
For more details on these terms, we refer the reader to Ref.~\onlinecite{kaga2022eliashberg}.
In the last term, we introduce \(\mathscr{H}_{\mathrm{ex}}\), which represents a test field of strength $\alpha$ lowering the symmetry. An explicit definition of this term is provided below.

We take $\omega_\eta=\omega_0=0.15$ and $g_\eta=g_0=\sqrt{3\lambda_0\omega_0/4}$ with $\lambda_0=0.15$. These parameters are adopted from Ref.~\onlinecite{kaga2022eliashberg}.
We set $U=J=\mu=0$ and consider only the electron-phonon interaction to investigate low-rank structures hidden in Feynmann diagrams functions. 
Numerical treatment of the instantaneous part will be more straightforward.

\begin{figure}
    \centering
    \includegraphics[width=1.0\linewidth]{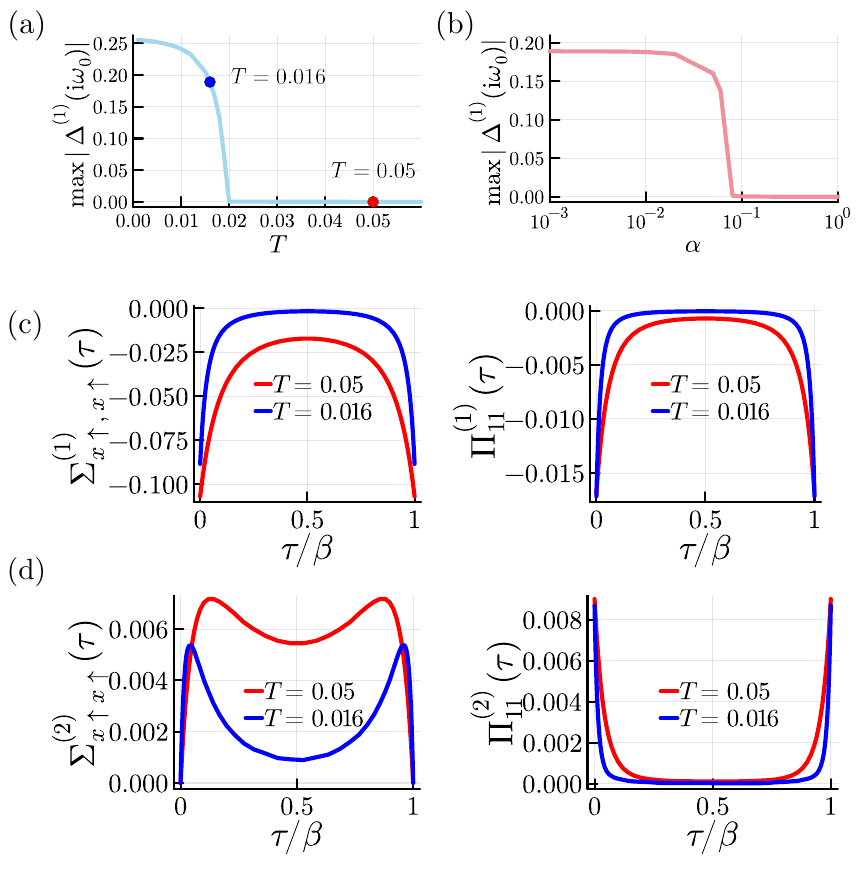}
    \caption{
    (a) $T$ dependence of the absolute maximum of the gap function with no external field ($\alpha=0$).
    (b) $\alpha$ dependence of the absolute maximum of the gap function at $T=0.016$.
    (c) $\Sigma^{(1)}$ and $\Pi^{(1)}$ computed at $T=0.05,~0.016$ with $\alpha=0$.
    (d) $\Sigma^{(2)}$ and $\Pi^{(2)}$ computed at $T=0.05,~0.016$ with $\alpha=0$.
    }
    \label{fig:selfconsistent}
\end{figure}

We solve the model within the dynamical-mean field theory (DMFT) by assuming local self-energies.
The local Green's function in the Nambu basis reads
\begin{align}
    \check{G}_{\gamma\sigma,\gamma'\sigma'}(\tau) &= -\ev{\mathcal{T}\Psi(\tau)\Psi^\dag(0)}
    = 
    \left( \begin{smallmatrix} G_{\gamma\sigma, \gamma'\sigma'}(\tau) & F_{\gamma\sigma, \gamma'\sigma'}(\tau)\\ \bar{F}_{\gamma\sigma, \gamma'\sigma'}(\tau) & \bar{G}_{\gamma\sigma, \gamma'\sigma'}(\tau) \end{smallmatrix} \right),\nonumber\\
    D_{\eta\eta'}(\tau) &= -\ev{\mathcal{T}\phi_{\eta}(\tau)\phi_{\eta'}(0)},\nonumber
\end{align}
where
$\Psi(\tau) = (c_{x\uparrow}(\tau), c_{x\downarrow}(\tau), \cdots, c_{z\downarrow}(\tau), c^\dag_{x\uparrow}(\tau), \cdots, c^\dag_{z\downarrow}(\tau))^{\!\top}$, $c_{\gamma\sigma}(\tau)$/$c_{\gamma\sigma}^\dagger(\tau)$ represents an annihilation/creation operator for orbital $\gamma$ and spin $\sigma$ in the Heisenberg picture, $\mathcal{T}$ indicates time odering.
Note that matrices with a checkmark $\check{}$, such as $\check{A}$, are defined in the Nambu space and have a shape of $12\times 12 matrix$. 
One can promote a matrix $A$ to the Nambu space as follows:
$\check{A} =\mathcal{N}(A) \equiv \left( \begin{smallmatrix} A & 0\\ 0 & -A^{\!\top} \end{smallmatrix} \right)$.
The test field \(\mathscr{H}_{\mathrm{ex}}\) is defined in the matrix form as  \(\mathcal{N}(l_z \otimes \mathbf{1}_{2\times2}) (\equiv \check{\mathscr{H}}_{\mathrm{ex}})\) with $l_z = \left( \begin{smallmatrix} 0 & -\im & 0\\ \im & 0 & 0\\ 0 & 0 & 0 \end{smallmatrix} \right)$ being an orbital magnetic moment. 
The self-consistent equations within DMFT read
    \begin{align}
        \!\!\!\check{G}(\iw_n)\! &=\!\!
        \int\! d\varepsilon \rho(\varepsilon)\!\left[\im\omega_n \mathbf{1}_{12\times12} \!+ \!(\mu -\varepsilon)\check{\mathbf{1}}
        \!-\! \check{\Sigma}(\im\omega_n) \!-\!\alpha\check{\mathscr{H}}_{\mathrm{ex}}\right]^{-1},
        \nonumber\\
        D(\iv_m) &= D^{(0)}(\im\nu_m)[\mathbf{1}_{6\times 6} - D^{(0)}(\im\nu_m) \Pi(\im\nu_m)]^{-1},\nonumber
    \end{align}
    where $\omega_n$ and $\nu_m$ are Matsubara frequencies for fermions and bosons, 
    $\mathbf{1}_{N\times N}$ is the $N\times N$ identity matrix, $\check{\mathbf{1}} \equiv \mathcal{N}(\mathbf{1}_{6\times6})$
    and $\rho(\varepsilon) = (8/\pi)\sqrt{(1/2)^2-\varepsilon^2}$ 
    , $D_{\eta\eta'}^{(0)}(\iv_m) = 2\delta_{\eta\eta'}\omega_0/((\im \nu_m)^2 - \omega_0^2)$ is the non-interacting phonon Green's function. 
The electron's self-energy, denoted by $\check{\Sigma}$, is defined as a $12\times12$ matrix in the form of 
\begin{align}
    \check{\Sigma}(\tau) = 
    \begin{pmatrix}
        \Sigma(\tau) & \Delta(\tau)\\
        \Delta^\dag(\tau) & -\Sigma^{\!\top}(-\tau)
    \end{pmatrix}
    \equiv
    \begin{pmatrix}
        \left(\check{\Sigma}\right)_{11} & \left(\check{\Sigma}\right)_{12} \\
        \left(\check{\Sigma}\right)_{21} & \left(\check{\Sigma}\right)_{22} \\
    \end{pmatrix},
    \label{eq:checkSigma}
\end{align}
where $\Delta(\tau)$, referred to as the gap function, is a $6\times 6$ matrix.
The phonon Green's function $D$ and the self-energy $\Pi$ are $6\times 6$ matrix.

A weak-coupling expansion yields the following connected skeleton diagrams\changed{~\cite{potthoff2004non}}  for the electron and phonon self-energies [Fig.~\ref{fig:highlight}(a)]. Here, we consider up to the second-order diagrams, and \(\check{\Sigma} = \check{\Sigma}^{(1)} + \check{\Sigma}^{(2)}\). The first-order and second-order self-energies are
\begin{align}
    &\check{\Sigma}^{(1)}_{\gamma\gamma'}(\tau) = -\sum_{\eta_1\eta_2}g_0^2 D_{\eta_1\eta_2}(\tau)\Big[\check{\lambda}^{\eta_1}\check{G}(\tau)\check{\lambda}^{\eta_2}\Big]_{\gamma\gamma'},
    \label{eqs:Sigma1}\\
    &\check{\Sigma}^{(2)}_{\gamma\gamma'}(\tau) = \int_{[0,\beta]^2}d\tau' d\tau''\sum_{\eta_1\eta_2\eta_3\eta_4}g_0^4D_{\eta_1\eta_3}(-\tau'')D_{\eta_2\eta_4}(\tau'-\tau)\nonumber\\
    &~~~\times\Big[\check{\lambda}^{\eta_4}\check{G}(\tau-\tau'')\check{\lambda}^{\eta_3}\check{G}(\tau''-\tau')\check{\lambda}^{\eta_2}\check{G}(\tau')\check{\lambda}^{\eta_1}\Big]_{\gamma\gamma'},\label{eqs:Sigma2}\\
    &\Pi^{(1)}_{\eta\eta'}(\tau) = \dfrac{1}{2}g_0^2\Tr\left[\check{G}(- \tau)\check{\lambda}^{\eta'}\check{G}(\tau)\check{\lambda}^{\eta}\right]
    ,
    \label{eqs:Pi1}\\
    &\Pi^{(2)}_{\eta\eta'}(\tau) = -\dfrac{1}{2}\int_{[0,\beta]^2}d\tau'd\tau'' \sum_{\eta_2\eta_4}g_0^4D_{\eta_2\eta_4}(\tau'-\tau'')\times\nonumber\\
    &~\hspace{-0.3em}{\rm Tr}\left[\check{G}(-\tau'')\check{\lambda}^{\eta_4}\check{G}(\tau''-\tau)\check{\lambda}^{\eta}\check{G}(\tau-\tau')\check{\lambda}^{\eta_2}\check{G}(\tau')\check{\lambda}^{\eta'}\right],\label{eqs:Pi2}
\end{align}
where $\beta=1/T$ denotes inverse temperature.
The Gell-Mann matrix in the Nambu space reads $\check{\lambda}^\eta = \mathcal{N}(\lambda^\eta \otimes \mathbf{1}_{2\times2})$,
where $\mathbf{1}_{2\times2}$ is the unit matrix in spin space ($2\times2$). 
Hereafter, we refer to the integrands for the normal and anomalous parts of the electron self-energy~\eqref{eqs:Sigma2} the phonon self-energy~\eqref{eqs:Pi2} as $f^\Sigma$, $f^\Delta$, and $f^\Pi$, respectively.

We perform self-consistent calculations with $\Sigma^{(1)}$ and $\Pi^{(1)}$ using the sparse modeling methods based on the intermediate representation (IR)~\cite{Shinaoka2022-xv, Li2020-kb, shinaoka2017compressing} to describe a superconducting state (see Supplemental Material~\cite{supplement} for more details).
The resultant Green's functions are stored in IR and can be evaluated at any $\tau$.
This first-order solution is used for the evaluation of the more complex second-order diagrams involving imaginary time integrals.

\textit{Quantics tensor train}
We start with introducing the quantics representation of a univariate function, denoted by $f(x): x \in [0,1) \rightarrow \mathbb{C}$.
The function can be ``tensorized''
by binary notation~\cite{oseledets2009approximation, khoromskij2011d, shinaoka2023multiscale}. 
In practice, we construct an equidistant grid on the interval $[0, 1)$:
\(x = (0.\sigma_{1}\sigma_{2}\dots \sigma_{\scR})_2 = \sum_{r=1}^{\scR} \sigma_{r}2^{-r}\),
where $\sigma_r \in \{0, 1\}$ 
corresponds to the scale $2^{-r}$.
\changed{$\cR$ must be chosen such that $2^{-\scR}$ is finer than the finest structure of the function.}
We then decompose the $\cR$-way tensor, $F_{\sigma_1, \dots, \sigma_\sscR} \equiv f((0.\sigma_1 \cdots \sigma_\scR)_2)$, as
\begin{align}
    &F_{\sigma_1, \sigma_2, \cdots, \sigma_{\sscR}}
    \approx \sum_{\alpha_1=1}^{\chi_1}\sum_{\alpha_2=1}^{\chi_2}\cdots\sum_{\alpha_{\sscR-1}=1}^{\chi_{\sscR-1}}F^{(1)}_{\sigma_1\alpha_1}F^{(2)}_{\alpha_1\sigma_2\alpha_2}\cdots F^{(\scR)}_{\alpha_{\sscR-1}\sigma_{\sscR}},
    \label{eqs:Low-rank_approximate_QTT}\\
&\hspace{2em}\raisebox{-0.5\height}{\includegraphics[width=0.4\linewidth]{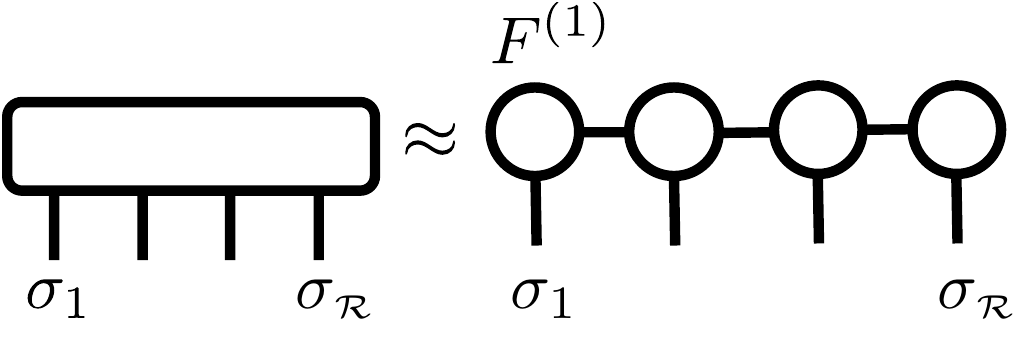}}~,
\end{align}
where $F_{\sigma_1, \dots, \sigma_\sscR}$ has a total of $2^\scR$ elements, and $F^{(\ell)}$ is a three-way tensor of size $2 \times \chi_{\ell-1} \times \chi_\ell$.
The sizes of the auxiliary indices (virtual bonds) $\alpha_{\ell}$, $\chi_\ell$, are called bond dimensions.
The bond dimension of the TT is defined as $\chi\equiv \mathrm{max}_\ell~\chi_\ell$.
One can integrate $f(x)$ accurately using Riemann summation 
by contracting each local index of dimension 2 with the vector \halfVector $\equiv (1/2, 1/2)$~\cite{supplement} as
\begin{align*}
\raisebox{-0.5\height}{
    \includegraphics[width=1.0\linewidth]{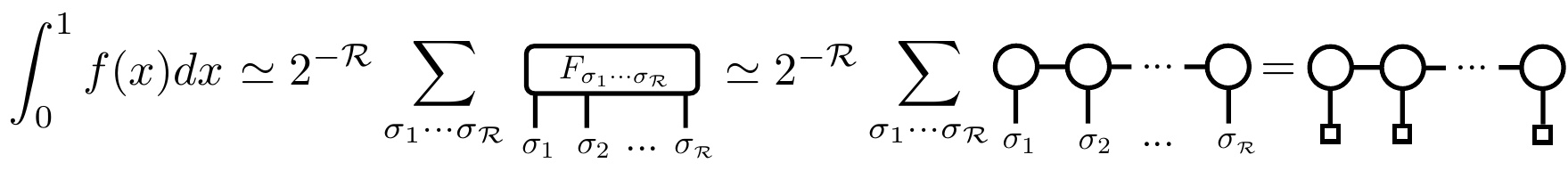}
    }~.
\end{align*}
In the above equation, the first $\simeq$ denotes the approximation by the Riemann sum, where the discretization error can be negligibly and exponentially small as $\mathcal{O}(2^{-\scR})$. The second $\simeq$ corresponds to the TT approximation. 
The computational time of the contraction scales linearly with $\cR$ as $\mathcal{O}(\chi^2 \cR)$.

The exact representation of an incompressible function, for example, a random function, requires $\chi_\ell$ to grow toward the center of the TT exponentially, offering no memory advantage over $F_{x_1, \dots, x_\sscR}$.
In the presence of a low-rank structure in $F$, i.e., the separation between different length scales~\cite{shinaoka2023multiscale}, the bond dimension can be significantly reduced; as a result, the total number of elements in the TT grows only linearly with $\cR$ as $\mathcal{O}(\chi^2 \cR)$.
The imaginary-time dependence of propagators self-energies is expected to have such low-rank structures~\cite{shinaoka2023multiscale,takahashi2024compactness}.

\textit{TT representations for integrands}
The integrands of self-energies (\ref{eqs:Sigma2}) and (\ref{eqs:Pi2}) depend on multiple discrete and continuous variables. For example, the integrand of the electron self-energy (\ref{eqs:Sigma2}) depends on discrete variables $\gamma, \gamma', \eta_1, \eta_2, \eta_3, \eta_4$, and continuous variables $\tau, \tau', \tau''$.
We apply the binary coding to the continuous variables as
$\tau/\beta = (0.\tau_{1} \cdots \tau_{\scR})_2$,
$\tau'/\beta = (0.\tau'_{1} \cdots \tau'_{\scR})_2$,
$\tau''/\beta = (0.\tau''_{1} \cdots \tau''_{\scR})_2$
\changed{with $\cR = 20$ to ensure negligible discretization error~\footnote{
    \changed{The minimum number of bits $\cR_\mathrm{min}$ scales as $\cR_\mathrm{min} \propto \log \beta$.}
}.}
Figure~\ref{fig:highlight}(a) illustrates the TTs for the integrands.
The lengths of the TTs are $\cL = 3\cR + 6$ and $\cL = 3\cR + 4$, respectively.
We assign $\tau_r$, $\tau'_r$, $\tau''_r$, at the same length scale to the same tensor due to the strong entanglement between these degrees of freedom~\cite{shinaoka2023multiscale,takahashi2024compactness}.
The indices for the discrete variables are grouped on the left side of the TTs.

\textit{Integration over internal variables}  
    Once the TT representation is constructed, summation and integration over the internal variables can be performed efficiently, as explained above.
    For instance, the TT representation of the electron self-energy \eqref{eqs:Sigma2} can be constructed as~\footnote{
    We must include the factor $\beta^2$ due to changing variables: $x'=\tau'/\beta$ and $x''=\tau''/\beta$.}
    \begin{align*}
    \raisebox{-0.5\height}{
        \includegraphics[width=0.85\linewidth]{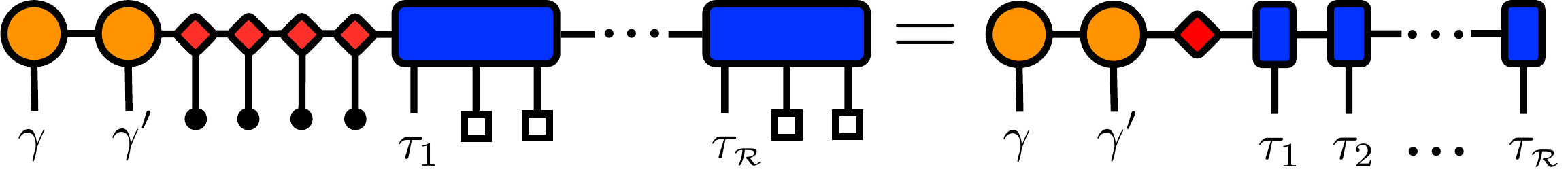}},
    \end{align*}
    where \onesVector is a vector filled with 1~\cite{supplement} and the tensor with no leg (red) can be absorbed into either its left (for $\gamma'$) or right (for $\tau_1$) tensor.
    Supplemental Material~\cite{supplement} provides a more detailed description.
    The resultant TT can be evaluated at any point $(\gamma, \gamma', \tau)$ with exponentially high resolution for $\tau$.

\textit{Tensor cross interpolation (TCI)} We construct a TT for the integrand by TCI, which is an adaptive learning algorithm that
constructs a low-rank TT from a small subset of adaptively chosen elements of a tensor $T$ (interpolation points).
Specifically, it optimizes $\chi_\ell$ so that a global error estimate satisfies $\|T - \tilde{T}\|_\mathrm{max}/\|T\|_\mathrm{max} < \epsilon$, where $\epsilon$ is a tolerance, $\tilde{T}$ is a TT and $\|\cdots \|_\mathrm{max}$ denotes the maximum norm.
Generally, the number of interpolation points required for constructing a TT with length $\mathcal{L}$ and bond dimension $\chi$ scales linearly with $\cL$ as $\mathcal{O}(\chi^2\mathcal{L})$, leading to an exponential speed-up over the conventional tensor decomposition based on singular value decomposition.
For details, please refer to Ref.~\onlinecite{fernandez2024learning}. 
\changed{We found that local 2-site updates in Ref.~\onlinecite{fernandez2024learning} face an ``ergodicity problem'' when exploring discrete variables. To overcome this problem, we added ``global updates'', combining greedy pivot optimization with random initial points, efficiently exploring variables with minimal overhead~\cite{supplement}.}

For constructing a TT for the integrand, we evaluate the RHS of Eq.~\eqref{eqs:Sigma2} and Eq.~\eqref{eqs:Pi2} for indices chosen by TCI.
This can be performed straightforwardly if the $\tau$ dependence of the Green's functions is expanded by IR.
The function evaluations consume the majority of the computational time of the TT construction.

We first discuss the $\alpha$-$T$ phase diagram.
Figure~\ref{fig:selfconsistent}(a) plots the $T$ dependence of the order parameter of the superconducting transition.
The system exhibits a second-order phase transition into a superconducting phase around $T=0.02$, 
being consistent with the results in the previous study~\cite{kaga2022eliashberg}.
Figure~\ref{fig:selfconsistent}(b) shows the $\alpha$ dependence of the order parameter at $T=0.016$. 
The order parameter vanishes around $\alpha=0.08$, signaling a transition to the normal phase.
\begin{figure}
        \centering
        \includegraphics[width=0.95\linewidth]{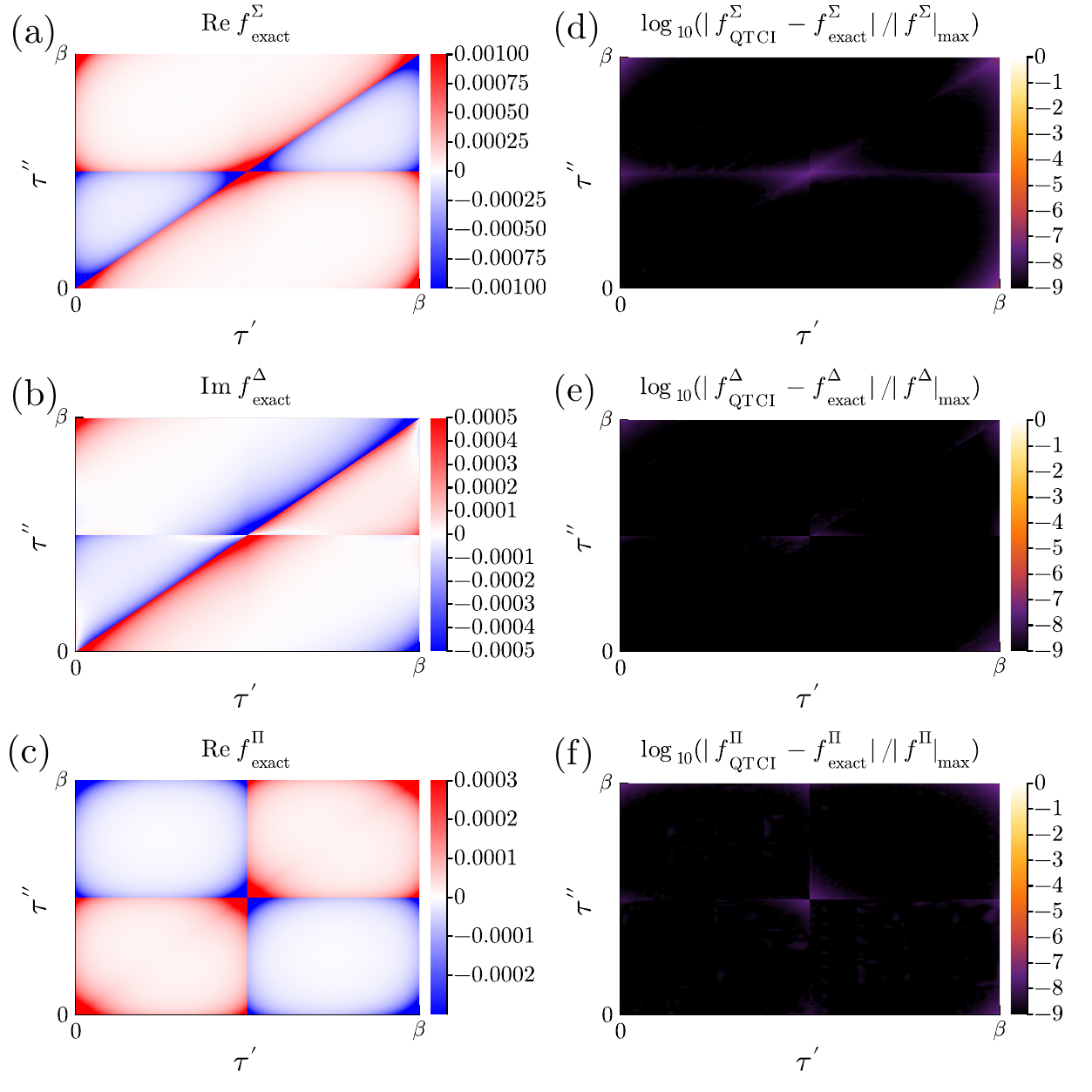}
        \caption{
        Integrands of (a) $\Sigma^{(2)}_{x\uparrow, x\uparrow}$, (b) $\Delta_{x\uparrow, x\downarrow}^{(2)}$, and (c) $\Pi^{(2)}_{11}$ at $\tau=\beta/2$ and $T=0.016$.
        Panels (d)--(f) show interpolation errors with $\epsilon=10^{-8}$ for the integrands in (a)--(c), respectively.
        }
        \label{fig:cut}
\end{figure}

Figure~\ref{fig:selfconsistent}(c) shows typical self-energies $\Sigma^{(1)}$ and $\Pi^{(1)}$ computed at $T=0.05$ ($>T_\mathrm{c}$) and $T=0.016$ ($< T_\mathrm{c}$) for $\alpha=0$.
Let us discuss the structures of the self-energy matrices in the flavor space.
At $\alpha=0$ and $T>T_\mathrm{c}$,
the normal part of the electron self-energy $\Sigma^{(1)}(\tau)$ and the phonon self-energy $\Pi^{(1)}(\tau)$ are diagonal, while the anomalous part of the electron self-energy $\Delta^{(1)}(\tau)=0$.
At $\alpha=0$ and for $T<T_\mathrm{c}$, the spontaneous symmetry breaking gives rise to nonzero superdiagonal and subdiagonal elements in $\Delta^{(1)}(\tau)$.
Introducing $\alpha>0$ further reduces the symmetry.
We observed that  nonzero components appear in $(\gamma\sigma, \gamma'\sigma')=(x\!\!\uparrow, y\!\!\uparrow), (x\!\!\downarrow, y\!\!\downarrow)$ for $\Sigma$, $(\gamma\sigma, \gamma'\sigma')=(x\!\!\uparrow, y\!\!\downarrow), (x\!\!\downarrow, y\!\!\uparrow)$ for $\Delta$, and $(\eta, \eta')=(1,3), (4,6), (0, 8)$ for $\Pi$~\footnote{
For the other components, the integrands themselves can be nonzero but are integrated to 0. The integrand of $\Pi_{11}^{22}$ shown in Fig.~\ref{fig:cut}(b) is a typical example.
}, respectively.
This refers to the upper triangular part, and nonzero components also appear symmetrically in the lower triangular part.
The appearance of these nonzero components is consistent with what is expected from the symmetry of the external field. 
The second-order self-energy matrices have precisely the same structures of nonzero elements as those of the first-order self-energies.

The left panels of Fig.~\ref{fig:cut} show typical components of the integrands for $\Sigma^{(2)}$, $\Delta^{(2)}$, and $\Pi^{(2)}$ computed at $\alpha=0$ and $T=0.016$.
These integrands have several lines of discontinuities, originating from the discontinuity of the one-particle propagators at $\tau = n\beta$ ($n \in \mathbb{Z}$).
\changed{Note that quantics representation does not require splitting the domain into multiple continuous regions, as in the non-quantics TCI approaches~\cite{fernandez2022learning,Erpenbeck2023-dz}.}

We now construct TTs of these functions for typical parameter sets of $T$ and $\alpha$ by TCI.
The tolerance is set to $\epsilon=10^{-3}$. 
The TCI construction takes a few minutes with one Apple M2 CPU core for $\chi\simeq 100$.

The results are shown in Fig.~\ref{fig:highlight}(b).
We numerically confirmed that the constructed $\tau$-dependent $12\times 12$ matrix $\check{\Sigma}$ satisfies the expected relation between the blocks: $\left(\check{\Sigma}\right)_{11}(\tau) = -\left(\check{\Sigma}\right)_{22}^{\!\top}(-\tau)$ and $\left(\check{\Sigma}\right)_{12}(\tau) = \left(\check{\Sigma}\right)_{21}^\dag(\tau)$ .
This indicates the stability of the TCI.

Figure~\ref{fig:highlight}(b) shows the bond dimensions $\chi_\ell$ computed for typical parameter sets.
The bond dimensions are significantly smaller than the worst (incompressible) case and strongly depend on $\ell$.
For the electron self-energy, $\chi_\ell$ takes large values at $\ell = 4$ (between $\eta_2$ and $\eta_3$) and at $\ell \simeq 8$.
The bond $\ell = 8$ connects the time scales of $\tau/\beta = 2^{-2}$ ($r=2$) and $2^{-3}$ ($r=3$).
For shorter time scales, $\chi_\ell$ monotonically decreases, allowing us to increase the time resolution exponentially.
We observed a similar trend for the phonon self-energy: $\chi_\ell$ takes large values around $\ell = 6$, corresponding to the bond between $r=2$ and $r=3$.

We move on to a more detailed discussion on the compactness of the quantics representation.
Figure~\ref{fig:highlight}(c) shows the $\chi$ dependence of an estimate of the interpolation error.
The interpolation error for all the parameter sets eventually decreases exponentially with $\chi$.
Let us first focus on the cases with no external field ($\alpha=0$) in more depth.
Comparing the results for $T=0.05~(>T_\mathrm{c})$ and $T=0.016~(<T_\mathrm{c})$, one can see that the symmetry breaking roughly doubles $\chi$ required to reach $\epsilon=10^{-2}$ from $\chi \simeq 35$ to $\chi \simeq 55$.
This can be attributed to the emergence of the nonzero components in $\check{\Sigma}$.

At $T=0.016$, increasing $\alpha=0$ to $\alpha=0.05$ in the superconducting phase does not change the behavior of the interpolation error above $\epsilon=10^{-2}$ qualitatively but leads to the emergence of a more slowly decaying tail below $\epsilon=10^{-2}$.
This tail may originate from the requirement of interpolating the small off-diagonal components in $\check{\Sigma}$ induced by the external field.
Further increasing $\alpha$ beyond the transition point, $\chi$ becomes substantially smaller with the disappearance of the spontaneous symmetry breaking.

The right panels of Fig.~\ref{fig:cut} show the interpolation error in the reconstructed self-energy for a typical parameter set, $T=0.016$ and $\alpha=0$.
One can clearly see that the interpolation does not suffer from the discontinuity.
The interpolation error has block-like structures, which are consistent with the binary coding.


We now demonstrate the integration of internal variables in the TT format.
Once we obtain the integrated TT, we can readily evaluate it at any $\tau$ and flavor pair.
Figure~\ref{fig:selfconsistent}(d) shows the self-energy evaluated on a dense grid for the typical components.
Note that $\Sigma^{(2)}$ and $\Pi^{(2)}$ are much smaller than $\Sigma^{(1)}$ and $\Pi^{(1)}$ in amplitude, justifying the validity of the one-shot calculation of $\Sigma^{(2)}$ and $\Pi^{(2)}$.
The $\tau$ dependence of $\Sigma^{(2)}$ and $\Pi^{(2)}$ does not exhibit any signature of the discretization due to the exponential resolution.
We further tested the correctness of the evaluated self-energies by projecting them onto the IR basis: We confirmed that the expansion coefficient decays as fast as expected (not shown).

\changed{In this letter, we explored low-rank tensor trains (TTs) for integrands in self-energy Feynman diagrams using a multiorbital electron-phonon model. We examined second-order diagrams for electron and phonon self-energies, discretizing the imaginary time with quantics representation and embedding discrete variables like orbitals and vibrational modes. The integrands' low-rank structures enable efficient compression and faster-than-power-law interpolation error convergence (Fig.~\ref{fig:highlight}), unaffected by symmetry breaking or discontinuities of the integrands without the need for domain partitioning in traditional methods.}


\changed{Future work could include electron-electron interactions in the electron-phonon model. 
Various optimizations are feasible: sparse grids for external imaginary-time variables~\cite{Shinaoka2022-xv, Li2020-kb,shinaoka2017compressing, Kaye2022-ad}, 
optimizing index order~\cite{Hikihara2023-mv} (see preliminary results),
combining global search with simulated annealing.
While the results of this study are promising, further research is needed to achieve faster-than-power-law convergence in higher-order expansions.
Extending this approach to replace QMC sampling at higher orders and developing hybrid methods with QTCI and specialized algorithms, e.g., Ref.~\onlinecite{kaye2024decomposing}, are potential research avenues. }

\begin{acknowledgments}
H.S. was supported by JSPS KAKENHI Grants No. 21H01041, No. 21H01003, No. 23H03817 and No. 22KK0226 as well as  JST FOREST Grant No. JPMJFR2232, Japan.
H.I. and H.S. were supported by JST FOREST Grant No. JPMJFR2232, Japan.
N.O and S.H were supported by 
JSPS KAKENHI Grants No. 21K03459 and No. 23H01130.
H.S. Thanks, Mark Ritter, Jan von Delft, Markus Wallerberger, and Jason Kaye, for the fruitful discussions.
We used \texttt{SparseIR.jl}~\cite{wallerberger2023sparse} for self-consistent calculations on the sparse frequency and time grids.
\end{acknowledgments}

\bibliography{main_revise2.bib}


\clearpage

\supplement

\include{supplementalmaterial_body}

\end{document}

%% file: supplementalmaterial_body.tex
In this Supplemental Material, we detail the TCI algorithms, the first-order self-consistent calculations, and the tensor train format integration.
We also show preliminary results on the dependence of the bond dimension on the ordering of TT indices, which is not detailed in the main text.

\section{Tensor cross interpolation with global searches}
\label{sec:tci}
To construct TCIs in this study, we combine the 2-site and global updates.
The 2-site updates can optimize the bond dimension by updating the TT through local searches, but they suffer from the so-called ergodicity problem discussed below.
Therefore, combining it with global searches that probabilistically explore the entire space allows us to explore regions the 2-site updates fail to sample.
In this section, we will explain the global update algorithm.

\subsection{Basic TCI algorithm}
\subsubsection{TCI formula}
The TCI allows the construction of an approximate tensor train by sampling only a small subset of the original tensor's elements, called \textit{pivots}.

We briefly explain how to construct a tensor train $\tilde{F}_{\bm{\sigma}}$ from the original $\cL$-leg tensor $F_{\bm{\sigma}}$.
We refer the reader to Ref.~\onlinecite{fernandez2024learning} for a more detailed and self-contained description.
We first define a multi-index, $\bm{\sigma} = (\sigma_1, \sigma_2, \cdots, \sigma_\scL)$, where $\sigma_\ell \in \mathbb{S}_\ell = \{1, \cdots, d_\ell\}$.
For the quantics representation, $d_\ell = 2$.
Then, we consider some subsets of multi-indices $\mathcal{I}_\ell\subseteq \mathbb{S}_1\otimes\mathbb{S}_2\otimes\cdots\otimes\mathbb{S}_\ell$ and $\mathcal{J}_\ell\subseteq\mathbb{S}_{\ell+1}\otimes\mathbb{S}_{\ell+2}\otimes\cdots\otimes\mathbb{S}_\scL$ for $\ell=1,2\cdots,\mathcal{L}-1$.

For given \(\{\mathcal{I}_\ell\}\) and \(\{\mathcal{J}_\ell\}\),
the TCI formula is given by
\begin{widetext}
\begin{align}
    &F_{\bm{\sigma}} \simeq \tilde{F}_{\bm{\sigma}} = [T_1]^{\sigma_1}_{i_0j_2}[P_1^{-1}]_{j_2i_1}[T_2]^{\sigma_2}_{i_1j_3}[P_2^{-1}]_{j_3i_2}\cdots [P_{\scL}^{-1}]_{j_\sscL i_{\sscL-1}}[T_\scL]^{\sigma_\sscL}_{i_{\sscL-1}j_{\sscL+1}},\label{eq:TCI}\\
    &\raisebox{-0.5\height}{
        \includegraphics[width=0.7\linewidth]{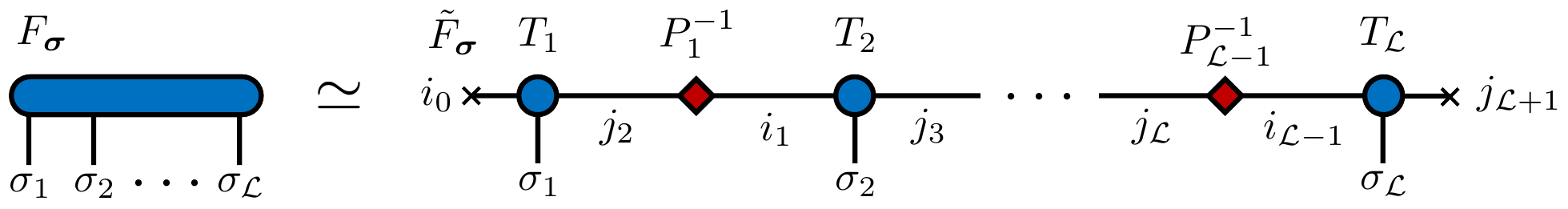}},
\end{align}
\vspace{-0.3cm}
\end{widetext}
where three-way tensors $T_\ell$ and matrices $P_\ell$ are defined by
$[T_\ell]^{\sigma_\ell}_{i_{\ell-1}j_{\ell+1}} = F_{i_{\ell-1}\oplus \sigma_\ell \oplus j_{\ell+1}}$ and $[P_\ell]_{j_{\ell+1}i_\ell} = F_{i_{\ell}\oplus j_{\ell+1}}$.
In Eq.~\eqref{eq:TCI}, we assumed Einstein summation over $i_\ell$, $j_\ell$ and introduced dummy indices fixed to 1.
The $T$ tensors and $P$ matrices are the slices of the original tensor $F_{\bm{\sigma}}$ defined by $\{\mathcal{I}_\ell\}$ and $\{\mathcal{J}_\ell\}$.
We call the selected subsets of multi-indices $\{\mathcal{I}_\ell\}$ and $\{\mathcal{J}_\ell\}$ \textit{pivots}.
The TCI~\eqref{eq:TCI} can be transformed to a TT by absorbing $P^{-1}$ into into $T$ tensors.
One can show that the resultant TT interpolates the original tensor on all pivots in the $T$ tensors and $P$ matrices if the pivots satisfy \textit{nesting conditions}.
For a more detailed discussion on the nesting conditions, we refer the reader to Sec.~4.2 in Ref.~\onlinecite{fernandez2024learning}.
If the original tensor is low-rank,
a few pivots approximate the whole original tensor.
Note that the sizes of the pivots determine the bond dimensions of the approximate TT.

\subsubsection{Local updates of pivots}
We use the 2-site update method based on partial rank-revealing LU decomposition (prrLU)~\cite{fernandez2022learning}.
We update pivots at $P_\ell$, i.e., $\mathcal{I}_\ell$ and $\mathcal{J}_{\ell+1}$ by computing the four-leg ``\(\Pi\)'' tensor 
for adjacent two local indices $\sigma_\ell$ and $\sigma_{\ell+1}$ and decompose it by prrLU for a given tolerance:
\begin{widetext}
\vspace{-0.5cm}
    \begin{align}
        [\Pi_\ell]_{i_{\ell-1}\sigma_\ell\sigma_{\ell+1}j_{\ell+2}} = F_{i_{\ell-1}\oplus \sigma_\ell \oplus \sigma_{\ell+1}\oplus j_{\ell+2}} \simeq [T^\prime_\ell]^{\sigma_\ell}_{i_{\ell-1}j_{\ell+1}}[P^{\prime -1}_\ell]_{j_{\ell+1}i_{\ell}}[T^\prime_{\ell+1}]_{i_\ell j_{\ell+2}}^{\sigma_{\ell+1}}.\\
        \raisebox{-0.5\height}{
        \includegraphics[width=0.6\linewidth]{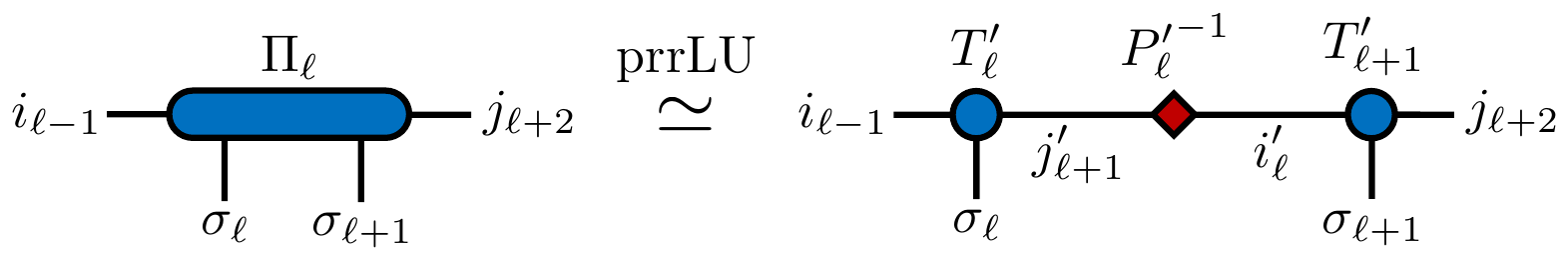}}.
\end{align}
\vspace{-0.3cm}
\end{widetext}
The prrLU selects relevant rows from \(\mathcal{I}_{\ell-1}\otimes \mathbb{S}_\ell\), and columns from \(\mathbb{S}_{\ell+1} \otimes \mathcal{J}_{\ell+2}\), respectively, through Gaussian elimination with full pivoting.
We replace $\mathcal{I}_{\ell}$ and $\mathcal{J}_{\ell+1}$ by the selected rows and columns.
We perform sweeps over all the pivot matrices until convergence.

This local algorithm, however, may fail to find some important features in the tensor, especially if the tensor is sparse or has discrete symmetries, such as those in spin and orbital sectors.
This leads to converge to an incorrect tensor train.
This issue is known as the ``ergodicity'' problem [see Sec.4.3.6 in \cite{fernandez2024learning}].
The ergodicity problem poses a serious challenge in systems with discrete degrees of freedom, such as orbitals and phonon vibrations, like those we dealt with in this study.

\subsection{Mitigating the ergodicity problem by global updates}
\begin{figure}
    \centering
    \includegraphics[width=0.65\linewidth]{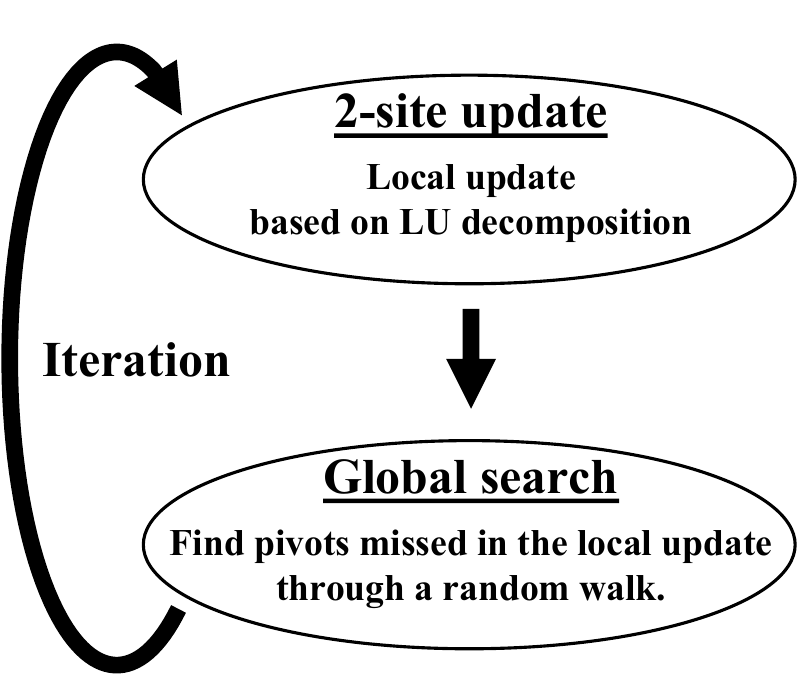}
    \caption{Schematic of the combination of local 2-site updates and global updates.}
    \label{fig:globalmove}
\end{figure}
\begin{figure}
    \centering
    \includegraphics[width=0.8\linewidth]{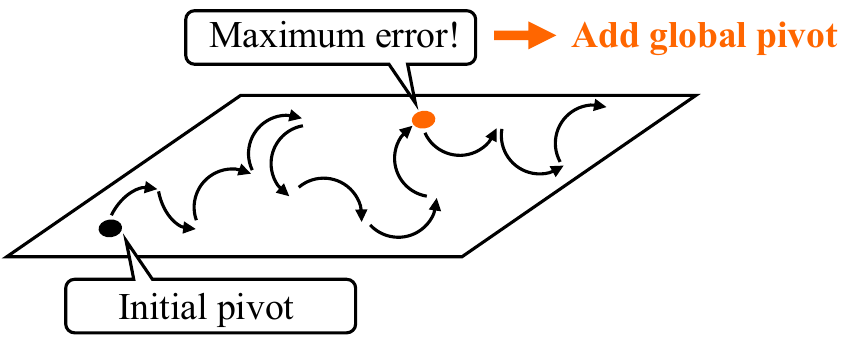}
    \caption{The global update randomly selects a pivot from the search space as a starting point and moves globally through the search space to find points with large interpolation errors.}
    \label{fig:globalupdates}
\end{figure}

To mitigate the ergodicity problem, we propose combining the 2-site updates and global updates, illustrated in Fig.~\ref{fig:globalmove}, \ref{fig:globalupdates} and described in Algorithm~\ref{alg:TCIalg}.
\changed{We iteratively perform 2-site updates and global updates. The global pivots are searched in a greedy manner across the entire search space with random initial points.}

In the global update part,
we perform global searches to identify multi-indices where the interpolation errors are still large [Algorithm~\ref{alg:globalsearches}].
The found \textit{global pivots} are then added to the pivot lists.
To be more specific, for a given global pivot $\tilde \sigma=\tilde \sigma_\ell \oplus \tilde \sigma_{\ell+1}$, we add $\tilde \sigma_\ell$, $\tilde \sigma_{\ell+1}$ to $I_{\ell}$, $J_{\ell}$, respectively, for all $\ell$.
For a detailed technical discussion on adding global pivots, we refer the reader to Sec.~4.3.5 in Ref.~\onlinecite{fernandez2024learning}.
\begin{algorithm}[H]
    \label{alg:TCIalg}
    \caption{TCI algorithm with 2-site updates + global updates \\ 
    The algorithm iterates over the 2-site local and global updates until the estimated TT's error meets the specified tolerance \(\epsilon\).
    }
    
    \KwIn{Function $F(\bm{\sigma})$, initial pivots \(\{\mathcal{I}_\ell\}\) and \(\{\mathcal{J}_\ell\}\), the number of maximum iteration $N_{\mathrm{max}}$, tolerance $\epsilon$}
    \KwOut{Tensor train $\tilde F$, pivots \(\{\mathcal{I}_\ell\}\) and \(\{\mathcal{J}_\ell\}\)}

    \For{i \KwFrom $1$ \KwTo $N_{\mathrm{max}}$}{
        perform a sweep with 2-site updates\
        
        compute tensor train $\tilde{F}$ from \(\{\mathcal{I}_\ell\}\) and \(\{\mathcal{J}_\ell\}\)

        search global pivots $\bm{P}$ by \textbf{Algorithm \ref{alg:globalsearches}}.\
        
        insert the global pivots to the pivot lists \(\{\mathcal{I}_\ell\}\) and \(\{\mathcal{J}_\ell\}\)
        
        \If{$|\bm{P}| = 0$ $\land$ $\|F - \tilde{F}\|_{\mathrm{max}}/\|F\|_{\mathrm{max}} < \epsilon$}{
        \KwBreak\
        }
    }
    
    \Return $\tilde{F}$, \(\{\mathcal{I}_\ell\}\),  \(\{\mathcal{J}_\ell\}\)
\end{algorithm}

\begin{algorithm}[H]
    \label{alg:globalsearches}
    \caption{Seaching a set of global pivots}
    \KwIn{The number of pivots searches $N_{\mathrm{search}}$, the number of pivot to be added $N_{\mathrm{pivot}}$, tolerance $\epsilon$}
    \KwOut{global pivots list $\bm{P}$}

    create empty global pivots list $\bm{P}$.
    
    \For{i \KwFrom $1$ \KwTo $N_{\mathrm{search}}$}{
        find a global pivot $\bm{p}$ with interpolation error $e$ by \textbf{Algorithm \ref{alg:globalsearch}}.
        
        \If{$e > \epsilon$}{
            $\bm{P} \gets \bm{P} \cup \{\bm{p}\}$
        }
    }
    \Return $\bm{P}$
\end{algorithm}

\begin{algorithm}[H]
    \label{alg:globalsearch}
    \caption{Global pivot search\\
    Starting from a randomly chosen multi-index, we perform sweeps over all local indices using a greedy algorithm to find a multi-index with a large interpolation error for a given tensor train approximation.
    }
    \KwIn{Tensor train approximation $\tilde{F}_{\bm{\sigma}}$, function $f(\bm{\sigma})$, maximum number of iteration $N_{\mathrm{iter}}$, tolerance $\epsilon$.}
    \KwOut{global pivot $\bm{p}$, error $e$.}

    generate initial multi-index $\bm{p} = (p_1, p_2, \cdots, p_\scL)$ randomly.

    initial error $e \leftarrow |f(\bm{p}) - \tilde{F}_{\bm{p}}|$ .
    
    \For{i \KwFrom $1$ \KwTo $N_{\mathrm{iter}}$}{
        \(e' \leftarrow e\)
        
        \For{j \KwFrom $1$ \KwTo $\cL$}{
            \(\bar{p}_j = \argmaxl_{p_j} |f(\bm{p}) - \tilde{F}_{\bm{p}}|\)
        
            \(\bm{p} \leftarrow (p_1, \cdots, p_{j-1}, \bar{p}_j, p_{j+1}, \cdots, p_\scL)\)
        
            $e = |f(\bm{p}) - \tilde{F}_{\bm{p}}|$

        }
        \If{$e = e'$}{
            \textbf{break}
        }
    }
    \Return $\bm{p},~ e$
\end{algorithm}

\subsection{Quantics TCI with/without the global updates}
\begin{figure}
    \centering
    \includegraphics[width=0.7\linewidth]{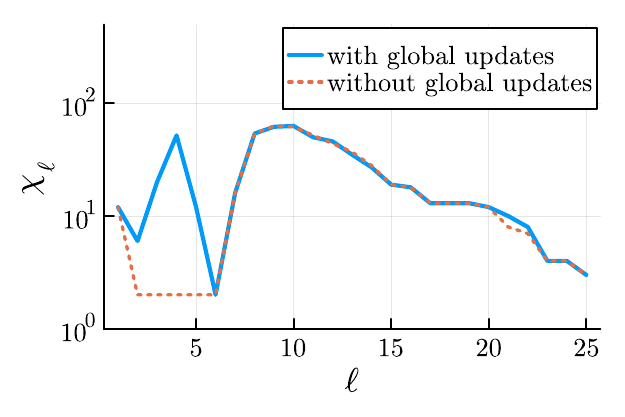}
    \caption{Comparison of bond dimensions with and without the global updates. Without the global updates, the TCI converges to the wrong result for \(\ell=2\) to \(\ell=6\).
    }
    \label{fig:bonddim_withwithout_global}
\end{figure}

We investigate the effectiveness of the global updates for the electron self-energy at \(T=0.016\).
Figure~\ref{fig:bonddim_withwithout_global} shows the bond dimensions
obtained with/without the global updates.
It can be seen that the bond dimensions are significantly smaller without the global updates for \(\ell=2\) to \(\ell=6\).
This part corresponds to the indices representing the electron's orbital and the phonon's vibrational modes.
This result indicates that the local updates failed to explore some of these sectors, and the resultant TT converged to the wrong one.

\begin{figure}
    \centering
    \includegraphics[width=1.0\linewidth]{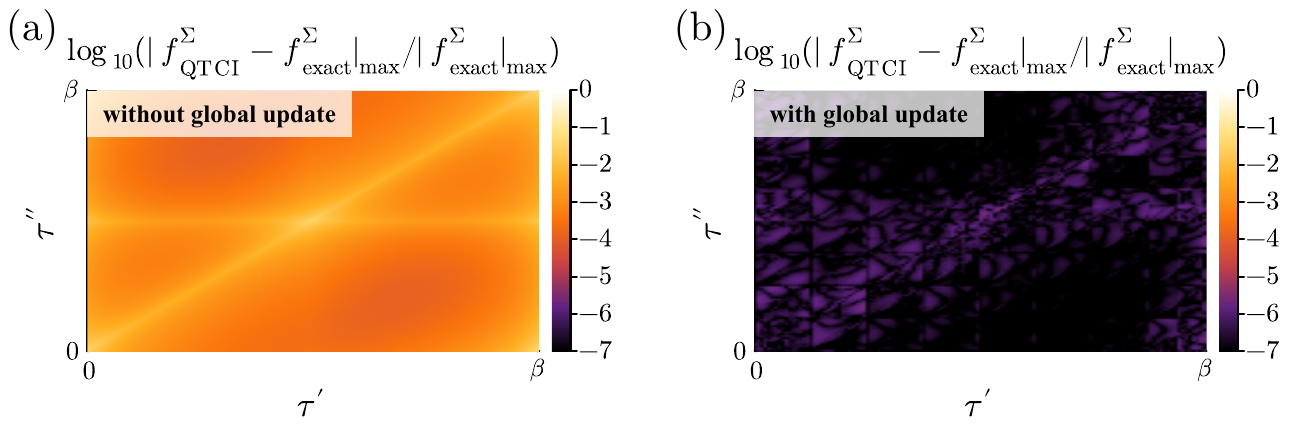}
    \caption{Two-dimension slice of the integrand of the self-energy on the plane of $(\tau', \tau'')$ at $(\gamma, \gamma', \eta_1, \eta_2, \eta_3, \eta_4, \tau) = (1,1,2,2,2,2,\beta/2), \beta=62.5, \alpha=0.0$.
    (a)Without global search and the exact value. (b)With global search and the exact value.}
    \label{fig:global_and_local_crosssection}
\end{figure}

\changed{Figure~\ref{fig:global_and_local_crosssection} shows the interpolation of the integrand of the electronic self-energy with and without global updates on a specific plane. 
Without global updates, the interpolation completely misses this section of discrete variables, while global updates successfully recover the missing part.
For the global updates, we used $N_{\mathrm{iter}}=N_{\mathrm{search}}=N_{\mathrm{max}} = 5$, which requires only minimal additional computational cost.
We also confirmed that the TCI with global updates consistently converged to the correct result across multiple trials.
Since global updates can be implemented with negligible computational overhead compared to the 2-site update, we recommend always including global updates in practice.
}

\section{Implementation of self-consistent calculations}
\label{sec:selfconsistent}
In this section, we detail our implementation of the self-consistent calculations based on the intermediate representation (IR) basis \cite{shinaoka2017compressing}.
The IR basis allows functions in quantum field theory, such as Green's function, to be represented on a sparse grid with very few points in Matsubara frequency and imaginary time.
Fourier transforms between Matsubara frequency and imaginary time can be performed using the IR basis, enabling closed quantum field theory calculations on a sparse grid.

In Sec.~\ref{sec:IR}, following~\cite{shinaoka2022efficient}, we introduce the definitions of the IR basis.
In Sec.~\ref{sec:selfconsistentloop}, we explain the implementation of the first-order self-consistent calculations. 
\subsection{Definition of IR basis}
\label{sec:IR}
Let us consider the one-particle Green's function in the imaginary time formalism:
\begin{align}
    G(\tau) = -\ev{\mathcal{T}_{\tau}c(\tau)c^\dag(0)},
\end{align}
where \(c\) and \(c^\dag\) are the annihilation and creation operators for particles, respectively, and \(\mathcal{T}_\tau\) denotes the time-ordering operator. The inverse temperature is denoted as \(\beta\), and the imaginary time \(\tau\) lies within the interval \([0, \beta]\). 
By performing a Fourier transform, the Green's function in Matsubara frequency is obtained as:
\begin{align}
    G(\im\omega_n) = \int_{0}^{\beta}d\tau e^{\im\omega_n\tau}G(\tau),
\end{align}
where \(\omega_n\) is a Matsubara frequency, given by \(\omega_m = (2m+1)\pi/\beta\) for fermions and \(\omega_m = 2m\pi/\beta\) for bosons.
After performing a Fourier transform to imaginary time, it becomes
\begin{align}
    G(\tau) = -\int_{-\omega_{\mathrm{max}}}^{\omega_{\mathrm{max}}}d\omega K(\tau,\omega)A(\omega), \quad (0<\tau<\beta),
\end{align}
where $K(\tau, \omega)$ is the analytic-continuation kernel given by
\begin{align}
    K(\tau,\omega) &= 
    \begin{cases}
        &\dfrac{e^{-\tau\omega}}{1 + e^{-\beta\omega}} ~~ (\text{Fermion}),\\
        &\dfrac{e^{-\tau\omega}}{1 - e^{-\beta\omega}} ~~ (\text{Boson}).
    \end{cases}
\end{align}
For bosons, the kernel diverges at \(\omega=0\). To address this, the spectral function is modified, and the Green's function is expressed as:
\begin{align}
    G(\tau) &\equiv -\int_{-\omega_{\mathrm{max}}}^{\omega_{\mathrm{max}}}d\omega K^\mathrm{L}(\tau,\omega)\rho(\omega),\nonumber\\
    \rho(\omega) &= 
    \begin{cases}
        A(\omega) ~~ & (\text{Fermion}),\\
        \dfrac{A(\omega)}{\tanh(\beta\omega/2)} ~~ & (\text{Boson}),
    \end{cases}
\end{align}
where $K^{\mathrm{L}} (\equiv (e^{-\tau\omega})/(1 + e^{-\beta\omega}))$ is the logistic kernel.
We expand the kernel as
\begin{align}
    K^{\mathrm{L}}(\tau,\omega) = \sum_{\ell=0}^{\infty}U_\ell(\tau)S_\ell V_\ell(\omega),
\end{align}
where \(S_\ell\) are the non-negative singular values decaying exponentially, and \(U_\ell(\tau)\) and \(V_\ell(\omega)\) are the singular functions (IR basis functions).
The number of the singular values above a given cutoff $\epsilon$ scales as $O(\log (1/\epsilon) \log \wmax \beta)$.
The basis functions $U_l(\tau)$ can be used to expand $G(\tau)$ compactly.
The expansion coefficients are given by
\begin{align}
g_l = \int^\beta_0 d\tau U_l(\tau) G(\tau) = S_l \int_{-\wmax}^\wmax d\omega \rho(\omega) V_l(\omega),
\end{align}
and $g_l$ decays as fast as the singular values $S_l$.
The Green's function in the Matsubara frequency space can be represented as
\begin{align}
G(\im \omega_m) &= \sum_l g_l U_l(\im \omega_m),
\end{align}
where  \(U_l(\im \omega_m)\equiv \int_{0}^{\beta}d\tau e^{\im\omega_n\tau}U_l(\tau)\).

One can generate a sparse grid \(\{\bar \tau_k\}\) in \(\tau\) and a sparse grid \(\{\im \bar \omega_k\}\) in the Matsubara frequency space, associated with the IR basis~\cite{Li2020-kb}.
Note that the imaginary-time sparse grids are generated from the zeros of the IR basis functions, and thus, we use the same grid for fermionic and bosonic cases.
The number of grid points is equal to or only slightly larger than the size of the IR basis~\cite{li2020sparse}.
This allows us to (Fourier) transform numerical data between these sparse grids efficiently and in a stable way (see Fig.~5 in Ref.~\cite{shinaoka2022efficient}).

\subsection{Applications to self-consistent calculations}
\label{sec:selfconsistentloop}
Figure~\ref{fig:1stselfconst} illustrates the flow of the first-order self-consistent calculations.
Note that each of all the numerical objects is defined on the respective sparse grid.
The \(\Sigma^{(0)}, \Pi^{(0)}\) are initial random values of the self-energy.
Steps 1 to 4 are repeated until \(\Sigma\) and \(\Pi\) converge. 

Step 1 involves a Fourier transform from the Matsubara-frequency grid to the imaginary-time grid.
In Step 2, the self-energy is calculated at each point in the imaginary-time grid using Eq.~(3) and Eq.~(5) in from the main text. The Fourier transform in Step 3 is similar to that in Step 1. Finally, in Step 4, \(G(\im\omega_n), ~D(\im\nu_m)\) are computed at each point
in the Matsubara frequency grid using the Dyson equation.


\begin{figure}
    \centering
    \begin{tikzpicture}
        \draw (0,0)node{$\Sigma^{(0)}(\im\omega_n) ~~ \Pi^{(0)}(\im\nu_m)$};
        \draw (1.7,0)node[right]{randomize};
        
        \draw [->, >=stealth](0,-0.4)--(0,-1) node[right]{Step.0~ Dyson's equation (same as Step4)}--(0,-1.6);
        
        \draw (0,-2)node{$G(\im\omega_n)~~D(\im\nu_m)$};
        \draw [->, >=stealth](0,-2.4)--(0,-3)node[right]{Step.1~ Fourior transformation}--(0,-3.6);
        
        \draw (0,-4)node{$G(\tau)~~D(\tau)$};
        \draw [->, >=stealth](0,-4.4)--(0,-5)node[right]{Step.2~ perturbation}--(0,-5.6);
        
        \draw (0,-6)node{$\Sigma(\tau)~~\Pi(\tau)$};
        \draw [->, >=stealth](0,-6.4)--(0,-7)node[right]{Step.3~ Fourior transformation}--(0,-7.6);
        
        \draw (0,-8)node{$\Sigma(\im\omega_n)~~\Pi(\im\nu_m)$};
        
        \draw [->, >=stealth](0, -8.4)--(0,-10)node[align=left,right]{Step.4~ Dyson's equation\\~~~~~$\displaystyle G(\im\omega_n) = \int d\varepsilon\dfrac{\rho(\varepsilon)}{(\im\omega_n+\mu) - \varepsilon - \Sigma(\im\omega_n) - \alpha\mathscr{H}_{\mathrm{ex}}}$ \\ \\~~~~~$D(\im\nu_m)=\dfrac{D^{(0)}(\im\nu_m)}{1 - D^{(0)}(\im\nu_m)\Pi(\im\nu_m)}$}--(0, -11.6);
        
        \draw (0,-12)node{$G(\im\omega_n)~~D(\im\nu_m)$};
    \end{tikzpicture}
    \caption{Flow of self-consistent calculations 
    }
    \label{fig:1stselfconst}
\end{figure}


\section{Summation in the tensor train format}
\label{sec:sum_and_integration}
This section explains how to perform summation over indices in the tensor train format. Additionally, we describe how this method is used to perform Riemann integration using QTT.

\subsection{Cases of matrices}
\label{sec:summation}
For simplicity without loss of generality, we consider a tensor of degree 2, \(M_{i_1, i_2}\), where the dimensions of the indices are \(d_1\) and \(d_2\), respectively.
Taking the sum over the indices \(i_1\) and \(i_2\) means computing \(\sum_{i_1=1}^{d_1}\sum_{i_2=1}^{d_2}M_{i_1, i_2}\), as illustrated on the left side of Fig.~\ref{fig:matrix_tt_sum}(a).
This operation can be represented as a contraction between each index and a vector whose components are all 1.
In fact, if \(\bm{u}\) and \(\bm{v}\) are vectors of appropriate length with all components equal to 1, then \(\bm{u}^{\!\top}M\bm{v}\) is the sum of all elements.
In the diagram, the vectors are represented as the black dots \onesVector, as shown in the middle part of Fig.~\ref{fig:matrix_tt_sum}(a).
Here, the black dot on the left is a vector of length \(d_1\) with all components equal to 1, and the black dot on the right is a vector of length \(d_2\) with all components equal to 1. The equality connecting the left and middle parts is exact.
The same discussion applies to tensors of order three or higher.
In this way, the sum over the indices of a tensor can be performed.

\subsection{Cases of tensor trains}
\begin{figure}
    \centering
    \includegraphics[width=1.0\linewidth]{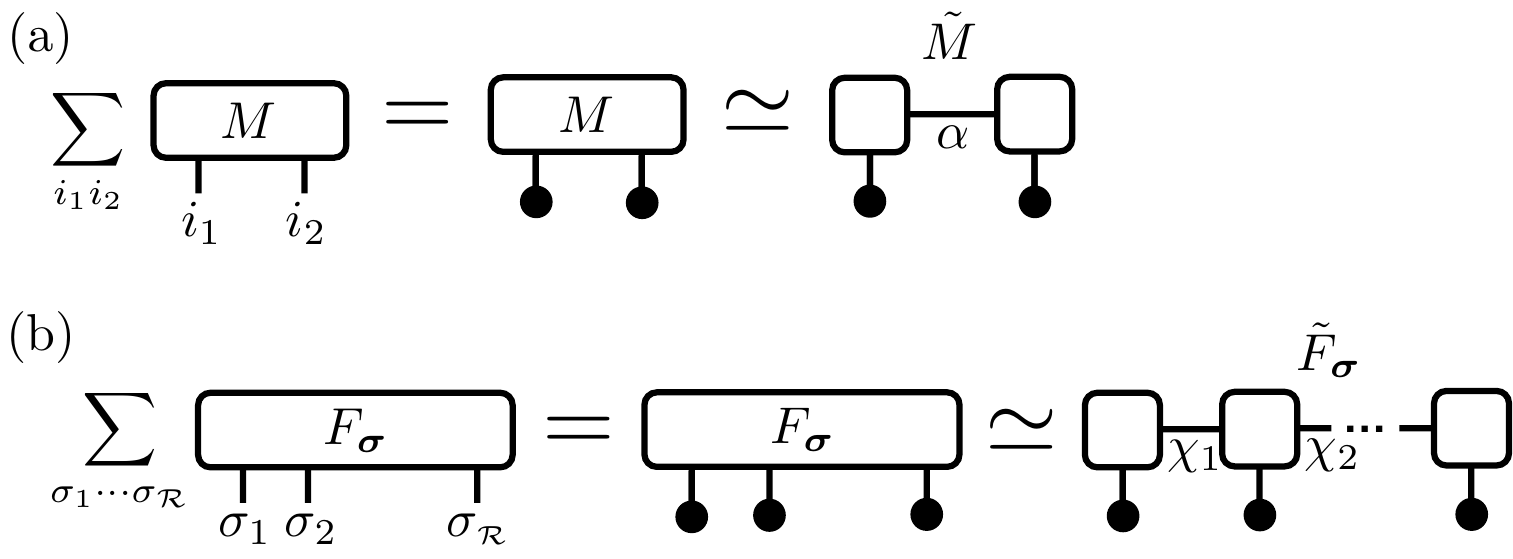}
    \caption{
    (a) Schematic diagram of summation over the matrix indices. (b) Schematic diagram of summation over the higher-order tensor indices.}
    \label{fig:matrix_tt_sum}
\end{figure}

Next, we consider a tensor train approximating tensor $F_{\bm{\sigma}}$.
Since the summation over local indices and the tensor-train decomposition are commutable, we obtain
\begin{align}
    & \sum_{\bm{\sigma}}F_{\bm{\sigma}}\nonumber\\
    &= \sum_{\sigma_1\sigma_2\cdots\sigma_\sscR}F_{\sigma_1\sigma_2\cdots\sigma_\sscR}\\
    &\simeq \sum_{\sigma_1\sigma_2\cdots\sigma_\sscR}\sum_{\chi_1\chi_2\cdots\chi_{\sscR-1}}F^{(1)}_{\sigma_1\chi_1}F^{(2)}_{\chi_1\sigma_2\chi_2} \cdots F^{(\sscR)}_{\chi_{\sscR-1}\sigma_\sscR}\\
    &= \sum_{\chi_1\chi_2\cdots\chi_{\sscR-1}}\sum_{\sigma_1\sigma_2\cdots\sigma_\sscR}F^{(1)}_{\sigma_1\chi_1}F^{(2)}_{\chi_1\sigma_2\chi_2} \cdots F^{(\sscR)}_{\chi_{\sscR-1}\sigma_\sscR}\\
    &= \sum_{\chi_1\chi_2\cdots\chi_{\sscR-1}}\tilde F^{(1)}_{\chi_1}\tilde F^{(2)}_{\chi_1\chi_2}\cdots \tilde F^{(\sscR)}_{\chi_{\sscR-1}},
\end{align}
where $F^{(r)}$ denotes the $r$-th core tensor of the tensor train and $\tilde F^{(r)} \equiv \sum_{\sigma_r} F^{(r)}$.
These equations are illustrated in Fig.~\ref{fig:matrix_tt_sum}(b).
The symbol $\simeq$ represents the error from the decomposition into the tensor train.
The diagram clearly shows that the summation can be performed locally by contracting each tensor core with \onesVector.

\vspace{3mm}

\section{Dependence of the bond dimensions on the order of indices}
When constructing tensor trains from a continuous function, we must choose a specific arrangement (order) of the indices in the TT.
For cases with only continuous variables, it is empirically known that arranging them in order of descending length scales results in small bond dimensions~\cite{shinaoka2023multiscale}.
However, the optimal order is not straightforward when both discrete and continuous variables are present.

In the case of the electron self-energy, there are three types of indices: the orbital \(\gamma\), the phonon vibrational mode \(\eta\), and the imaginary time \(\tau\).
In the main text, the analysis was conducted using a tensor train with the indices ordered as \((\gamma, \eta, \tau)\).
We examined how the bond dimension depends on six possible orders of these three types of variables.

\begin{figure}[H]
    \centering
    \includegraphics[width=1.0\linewidth]{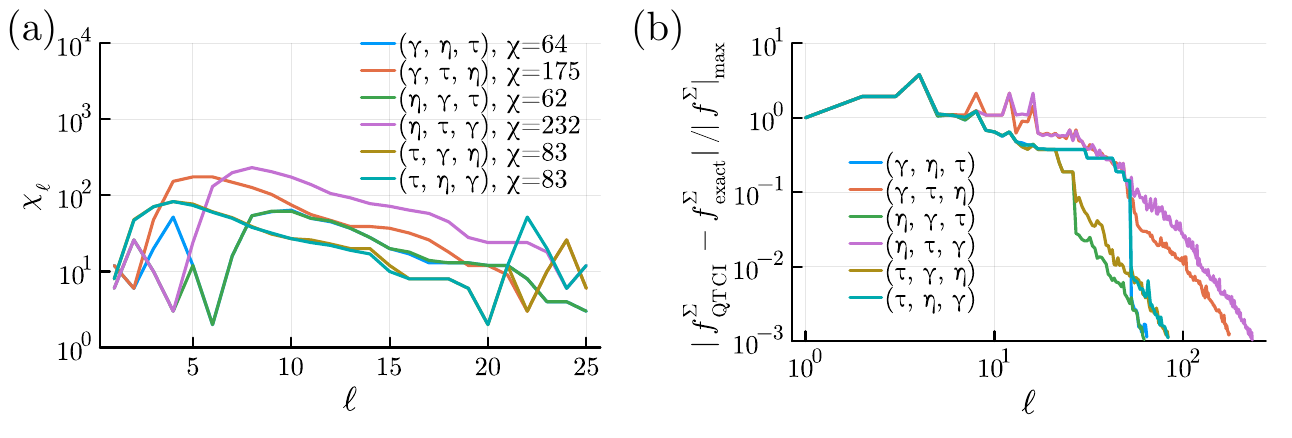}
    \caption{Dependence of the error convergence on the order of indices for the integrand of the electron self-energy at \(\alpha=0.0\) and \(T=0.016\).
    (a) The bond dimension of each order. Each interval for the respective variables exhibits roughly similar behavior. 
    (b) Interpolation error estimate as a function of bond dimension $\chi$. Although the convergence speed differs, the exponential convergence is robust.}
    \label{fig:order_bonddim_error}
\end{figure}

Figure~\ref{fig:order_bonddim_error} shows the convergence of bond dimensions and interpolation errors for the six index orders obtained by TCI.
The exponential decay of the error is robust regardless of the order. 
However, for the two index orders where the orbital \(\gamma\) and vibrational mode \(\eta\) are placed apart, the convergence becomes slower, and 2 to 3 times larger bond dimensions are required to reach the same tolerance compared to the other four orders.
This result suggests that when there are multiple types of discrete variables, it is more efficient to place them adjacent to each other. A more detailed investigation of this will be left for future work.

\section{Comparison of the Efficiency of QTCI and GKTCI}
When constructing a TT of a function, there is arbitrariness in the construction method.
When using TT for the integration of a function, in addition to the QTT introduced in this paper, a construction method using Gauss-Kronrod integration~(GK) can also be considered \cite{fernandez2022learning,Erpenbeck2023-dz,dolgov2020parallel}.
Here, we compare the efficiency of QTCI and TCI based on Gauss-Kronrod quadrature (GKTCI), where the Gauss-Legendre rule is used to construct the Gauss-Kronrod quadrature.

The variables of the integrand function for the second-order electron self-energy are explicitly written as
$f^{\Sigma}(\gamma, \gamma', \eta_1, \eta_2, \eta_3, \eta_4, \tau, \tau', \tau'')$.
This function exhibits discontinuities with respect to imaginary time.
In the QTT approach, the continuous variables $\tau$, $\tau'$, and $\tau''$ are discretized using an evenly spaced grid based on binary representation.
This exponentially refined grid allows QTT to effectively disregard the impact of discontinuities.
In contrast, in the Gauss-Kronrod (GK) approach, the integration variables $\tau'$ and $\tau''$ are discretized at the evaluation points of the Gauss-Legendre quadrature.
Since discontinuities cannot be neglected in this case, it is necessary to partition the domain into continuous regions.
The number of such regions increases as $\mathcal{O}((2n-1)!)$, where $n$ is the perturbation order.
For second-order perturbation ($n=2$), the discontinuities divide the domain into six regions.

In this work, the external variable $\tau$ is discretized using the quantics representation, while the internal variables $\tau'$ and $\tau''$ are discretized at the Gauss-Kronrod quadrature nodes.
We focus on the continuous region defined by $\tau' < \tau'' < \tau$, assuming no external field ($\alpha=0.0$).
The corresponding tensor-train (TT) diagram is shown below:
\begin{align*}
\includegraphics[width=0.9\linewidth]{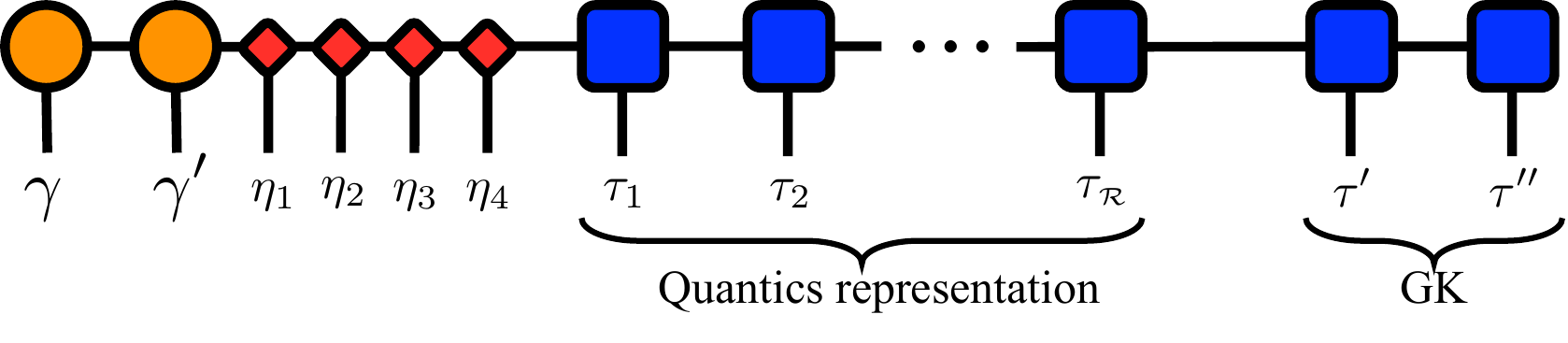}
\end{align*}
Here, the dimensions of the discretized variables $\tau'$ and $\tau''$ correspond to the number of Gauss-Kronrod quadrature nodes, $N_{\mathrm{GK}}$.

\begin{figure}[H]
    \centering
    \includegraphics[width=0.7\linewidth]{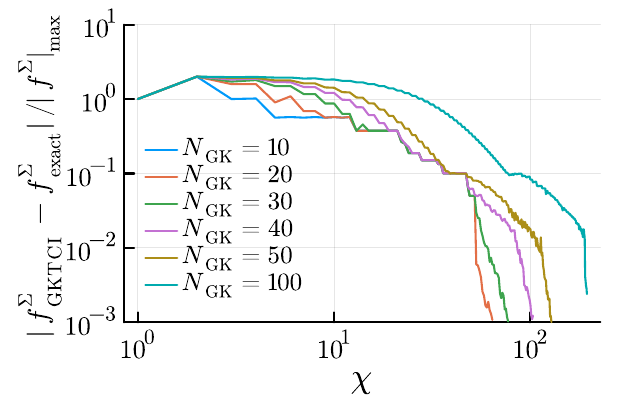}
    \caption{Interpolation error behavior when TCI is performed by discretizing the integration variables at the GK evaluation points. The continuous region $\tau' < \tau'' < \tau$ was selected, with $\beta=62.5$ and $\alpha=0.0$.}
    \label{fig:GK_error_decay}
\end{figure}

Figure~\ref{fig:GK_error_decay}(a) shows the error decay in the GKTCI method.
It is observed that the error decays exponentially regardless of the number of quadrature nodes, suggesting that GKTCI exhibits exponential convergence similar to QTCI when restricted to continuous regions.
However, even though the present results for GKTCI are computed for only a single continuous region, the computational cost was comparable to that of QTCI covering all regions.
This indicates that calculating each divided region as a separate TT may be inefficient from a computational standpoint.
Additionally, GKTCI requires careful tuning of more hyperparameters than QTCI, such as the number of quadrature points and the selection of pivot elements for large local dimensions.

